\documentclass{aa}
\usepackage{txfonts}
\usepackage[dvips]{graphicx}
\newcommand{\kms}   {km~s$^{-1}$}

\newcommand{\nho}   {NH$_3$\,(1,1)}       
\newcommand{\nhd}   {NH$_3$\,(2,2)} 
\def\nh{NH$_3$}      
\def\iras{IRAS~05173$-$0555}
\def\Vlsr{$V_{\rm LSR}$}

\def\juz{\mbox{$J$=1$\rightarrow$0}}
\def\jtt{\mbox{$J$=3$\rightarrow$2}}

\begin{document}


\title{Dissection of the protostellar envelope surrounding IRAS~05173$-$0555 in L1634}

\author{M.\ T.\ Beltr\'an\inst{1} \and J. Wiseman\inst{2} \and P.\ T.\ P.
Ho\inst{3, 4} \and R.\ Estalella\inst{1} \and G.\ A.\ Fuller\inst{5} \and A. Wootten\inst{6}}

\offprints{M. T. Beltr\'an, \email{mbeltran@am.ub.es}}

\institute{
Departament
d'Astronomia i Meteorologia, Universitat de Barcelona, Mart\'{\i} i Franqu\`es 1,
08028, Barcelona, Catalunya, Spain
\and
NASA-Goddard Space Flight Center, Greenbelt, MD 20771, USA 
\and
Harvard-Smithsonian Center for Astrophysics, 60 Garden Street, Cambridge, MA 02138, USA
\and 
Academia Sinica, Institute of Astronomy and Astrophysics, P.O.\ Box 23--141, Taipei 106, Taiwan
\and
Jodrell Bank Centre for Astrophysics, Alan Turing Building, University of Manchester, Manchester,
 M13 9PL, UK
\and 
National Radio Astronomy Observatory, 520 Edgemont Road, Charlottesville, VA 22903, USA}

\titlerunning{The protostellar envelope surrounding IRAS~05173$-$0555}
\authorrunning{Beltr\'an et al.}

\abstract
{The youngest protostars that power energetic outflows are surrounded by infalling and
rotating envelopes that contain most of the mass of the system.}
{To study the properties and kinematics of the protostellar envelope surrounding the embedded source \iras\  in L1634.}
{We carried out VLA ammonia observations at 1.3~cm  with the VLA in the D configuration to 
map the gas towards the core of L1634.}
{The \nh\ emission towards IRAS~05173$-$0555 is resolved and shows two components
clearly distinguishable morphologically: a cross-like structure, roughly
elongated in the direction of the HH~240/241 outflow and associated with \iras, plus an
arc-like stream elongated towards the north. The properties and kinematics of the gas suggest
that the origin of the cross-like morphology could be the interaction between the outflow and
the envelope. A more compact and flattened structure, which could be undergoing rotation about
the axis of the outflow, has been detected towards the center of the cross-like envelope. The northern stream, which has properties and velocity
different from those of the cross-like envelope, is likely part of the original cloud
envelope, and could be either a quiescent core that would never form stars, or be in a prestellar phase.}
{}

\maketitle

\section{Introduction}


The youngest protostars, which are associated with energetic molecular bipolar outflows, are
deeply embedded in circumstellar gas and dust material. Theory outlines a scenario where a
central object is surrounded by an infalling and rotating envelope that contains most of the
mass (e.g.\ Larson~\cite{larson69}; Adams et al.~\cite{adams87}; see Saigo et
al.~\cite{saigo08} and references therein for recent simulations of rotating and collapsing
cores). This infalling material is
accreted onto the central protostar funneled through a circumstellar disk that grows as the
system evolves (e.g.\ Shu et al.~\cite{shu87}). Hence, infalling, outflowing and rotation
motions take place simultaneously in such an extremely young environment, making the morphology
and kinematics of such regions very complex.

L1634 is a bright-rimmed cloud, SFO 16, associated with Barnard's Loop (Sugitani et
al.~\cite{sugitani91}; De Vries et al.~\cite{devries02}). It is a small, isolated dark cloud
located to the west of the Orion A cloud, at a distance of 460~pc (Bohigas et
al.~\cite{bohigas93}). The cloud has been proposed to be either a remnant of the molecular
material from which the nearby Orion OB1 association formed, or a cloud pushed to its present
location by the pressure associated with energetic events accompanying the evolution of the OB
association (Maddalena et al.~\cite{maddalena86}). Near-infrared observations show that the
cloud contains two H$_2$ bipolar jets (Hodapp \& Ladd~\cite{hodapp95}). One of these jets,
HH~240/241 (RNO 40), is a spectacular and powerful bipolar flow (Hodapp \&
Ladd~\cite{hodapp95}; Davis et al.~\cite{davis97}; Nisini et al.~\cite{nisini02}; O'Connell et
al.~\cite{oconnell04}) that extends in the east-west direction and is driven by
IRAS~05173$-$0555 ($L_{\rm bol}$$\simeq$17~$L_{\sun}$; Reipurth et al.~\cite{reipurth93}). This
molecular outflow has been mapped in CO by Lee et al.~(\cite{lee00}). The other bipolar jet
only has two knotty bow shocks (knots 9 and 4; Hodapp \& Ladd~\cite{hodapp95}), and the
near-infrared source found near the center, IRS~7 (Davis et al.~\cite{davis97}), is probably
the powering source of this second outflow. The continuum emission towards the core of L1634
has been mapped at centimeter, millimeter, and submillimeter wavelengths (Reipurth et
al.~\cite{reipurth93}; Chini et al.~\cite{chini97}; Beltr\'an et al.~\cite{beltran02}; Morgan
et al.~\cite{morgan08}). 

Beltr\'an et al.~(\cite{beltran02}) model the circumstellar dust emission around
IRAS~05173$-$0555, and conclude that the infrared source traces an embedded young stellar
object still in its infall phase. These authors classify the source as a Class~0 object, based
on the  infall rate estimate, the high ratio of submillimeter-to-bolometric luminosity, the
circumstellar mass estimates, the non-detectability of the source at near-infrared wavelengths
(Bohigas et al.~\cite{bohigas93}), and the fact that is driving a powerful outflow. However,
the source has been recently detected by the Infrared Array Camera (IRAC) of the {\it Spitzer Space
Telescope} at 3.6, 4.5, 5.8, and 8.0~$\mu$m, as shown by the Spitzer archive. Therefore, based
on the classification tests proposed by Froebrich~(\cite{froebrich05}), the source would be
better classify as a bordeline Class~0/I object, or maybe a Class~I object.

\begin{table*}
\caption[] {\nh\ line parameters$^{a}$}
\label{table_lines}
\begin{tabular}{llccccc}
\hline
 &
\multicolumn{1}{c}{Transition} &
\multicolumn{1}{c}{\Vlsr} &
\multicolumn{1}{c}{$\Delta V^{b}$}  &
\multicolumn{1}{c}{$A\tau_{\rm m}^c$} &
&\multicolumn{1}{c}{$T_{\rm L}^e$} 
\\
\multicolumn{1}{c}{Region} &
\multicolumn{1}{c}{($J, K$)} &
\multicolumn{1}{c}{(km s$^{-1}$)} &
\multicolumn{1}{c}{(km s$^{-1}$)} &
\multicolumn{1}{c}{(K)} &
\multicolumn{1}{c}{$\tau_{\rm m}^d$} & 
\multicolumn{1}{c}{(K)} 
\\
\hline
peak position  
&(1, 1) &$8.10\pm0.01$ &$0.51\pm0.01$ &$23.4\pm0.5\phantom{1}$ &$3.6\pm0.1$  &$6.0\pm0.2$ \\
&(2, 2) &$8.08\pm0.02$ &$0.52\pm0.05$ &$2.00\pm0.20$          &$0.31\pm0.03$ &$1.8\pm0.1$ \\
cross-like envelope  
&(1, 1) &$8.14\pm0.01$ &$0.52\pm0.01$ &$5.30\pm0.15$  &$2.3\pm0.1$   &$1.8\pm0.1$ \\ 
&(2, 2) &$8.08\pm0.02$ &$0.52\pm0.05$ &$0.59\pm0.09$ &$0.25\pm0.04$ &$0.45\pm0.06$ \\
northern stream  
&(1, 1) &$8.40\pm0.01$ &$0.51\pm0.01$ &$3.59\pm0.22$ &$2.1\pm0.3$ &$1.3\pm0.1$ \\ 
&(2, 2) &$\ldots^f$ &$\ldots^f$ &$<0.12\pm0.05$\phantom{22} &$<0.07\pm0.03$\phantom{22} &$<0.10\pm0.04^g$\phantom{8}  \\
\hline 
\end{tabular}

(a) Obtained from the fit, using the CLASS package, of the magnetic hyperfine components of the spectra (corrected
for the primary beam response), towards the emission peak position, averaged over an area
of $\sim$$1830$ arcsec$^2$
for the cross-like envelope, and $\sim$$1000$ arcsec$^2$ for the northern stream. \\
(b) Intrinsic line width (FWHM) of the magnetic hyperfine components. \\
(c) $A=f[J(T_{\rm ex})-J(T_{\rm bg})]$, where $f$ is the filling factor of the
emission, $T_{\rm ex}$ the
excitation temperature of the transition, $T_{\rm bg}$ the background radiation temperature and $J(T)$
the intensity in units of temperature, $J(T)=\frac{h\nu}{k}[exp(h\nu/kT)-1]^{-1}$. The filling factor has been assumed 
$f\simeq$1. \\  
(d) Optical depth of the main line. For the (1, 1) line, the optical depths have been
obtained from the hyperfine fit. For the (2, 2) line, the optical depths have been estimated
assuming that the filling factor and the excitation temperature are the same for both
transitions.  \\
(e) Main beam brightness temperature of the (1, 1) and (2, 2) main lines. \\
(f) NH$_3$(2,2) not detected. \\
(g) $3\,\sigma$ upper limit. 
\end{table*}

In this paper we present ammonia high-angular resolution observations towards the protostellar
envelope surrounding IRAS~05173$-$0555 in L1634 that improve our picture of its structure. The
data constrain the physical properties of the embedded IRAS~05173$-$0555 source and of its
surrounding envelope. In addition, the kinematical study of the gas allows us to infer the
presence of a rotating structure towards IRAS~05173$-$0555.

\section{Observations}

Interferometric observations were carried out in 2000 August 27 with the Very Large Array
(VLA) of the National Radio Astronomy Observatory (NRAO)\footnote{NRAO is a facility of the
National Science Foundation operated under cooperative agreement by Associated Universities,
Inc.} in the D configuration. The ammonia ({\it J, K}) = (1,1) and (2,2) inversion transitions
were observed simultaneously, at rest frequencies 23.694495~GHz and 23.722633~GHz,
respectively. The phase center of the observations was 
$\alpha(J2000)=5^\mathrm{h} 19^\mathrm{m} 49\fs0$, 
$\delta(J2000)=-5\degr 52' 0\farcs0$.
The width of the antenna primary beam response was approximately $2\arcmin$. The
resulting synthesized beam was $10\farcs4\times8\farcs0$ at P.A.\ = 6\degr, with natural
weighting and a 20~k$\lambda$ taper applied to the {\sl uv} data. The bandwidth was 3.125~MHz and
the channel separation was 24.4~kHz, corresponding to 0.31~\kms. Absolute flux calibration was
achieved by observing 3C~147, with an adopted flux density of 1.81~Jy at 1.3~cm. The bandpass
calibrator was 3C~48, and phase was calibrated by observing 0501$-$019, which has a
bootstrapped flux of 1.12$\pm$0.04~Jy. Reduction and analysis of the data were carried out
using standard procedures in the MIRIAD, AIPS, and GILDAS software packages.

\begin{figure*}
\centerline{\includegraphics[angle=-90,width=16cm]{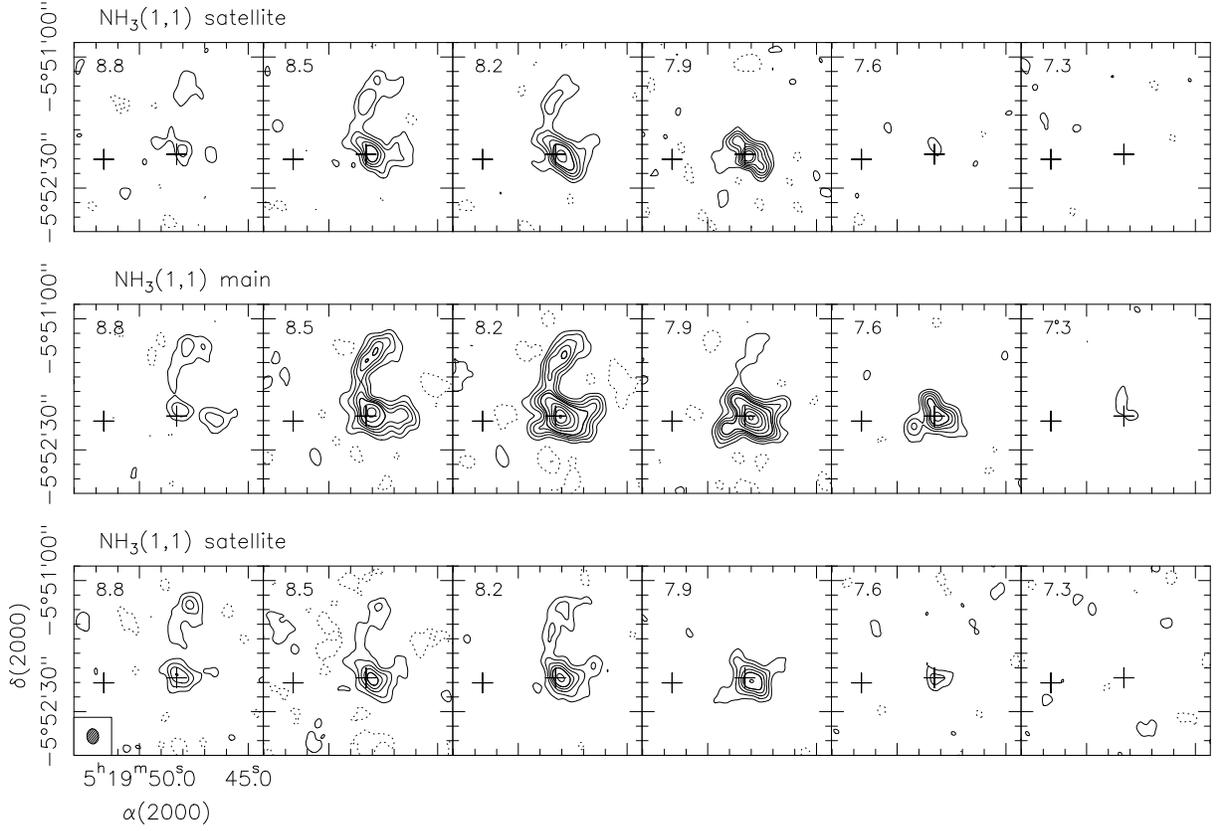}}
\caption{Velocity channel maps of the \nho\ emission for the inner satellite line
$\mbox{$F_1$=2$\rightarrow$1}$ ({\it top panel}), main line ({\it middle panel}), and  inner
satellite line $\mbox{$F_1$=1$\rightarrow$2}$ ({\it bottom panel}). The systemic velocity of L1634
is $8$~\kms.  The central velocity of each channel is indicated at the upper left corner of the
panels.  The $1\,\sigma$ noise in one channel is 5~mJy\,beam$^{-1}$. Contour levels are $-6$, $-3$,
3, 6, 9, 12, 15, 21, 27, 33, 39 and 45 times $\sigma$.  The conversion factor from Jy\,beam$^{-1}$
to K is 26.3. The synthesized beam is drawn in the bottom left  corner of the bottom left panel.
The crosses mark the submillimeter continuum positions of IRAS~05173$-$0555 ({\it right cross}) and
IRS~7 ({\it left cross}) (Beltr\'an et al.~\cite{beltran02}).} 
\label{nh311_channel}
\end{figure*}

\section{Results and analysis}

\subsection{Morphology of the NH$_3$ emission}

\begin{figure*}
\centerline{\includegraphics[angle=-90,width=14cm]{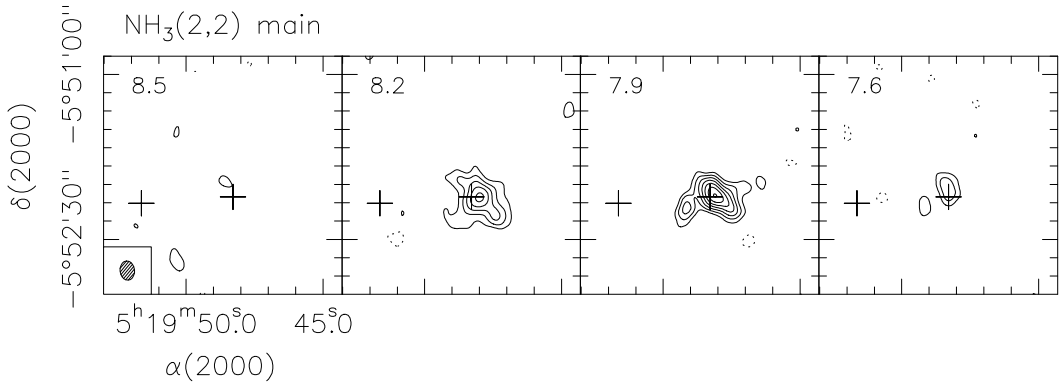}}
\caption{Same as Fig.~\ref{nh311_channel} for the \nhd\ emission. The $1\,\sigma$ noise in one channel is 4~mJy\,beam$^{-1}$.  Contour levels are $-3$,
3, 6, 9, 12, and 15 times $\sigma$. The conversion factor from
Jy\,beam$^{-1}$ to K is 26.2.} 
\label{nh322_channel}
\end{figure*}

\begin{figure}
\centerline{\includegraphics[angle=0,width=7.5cm]{{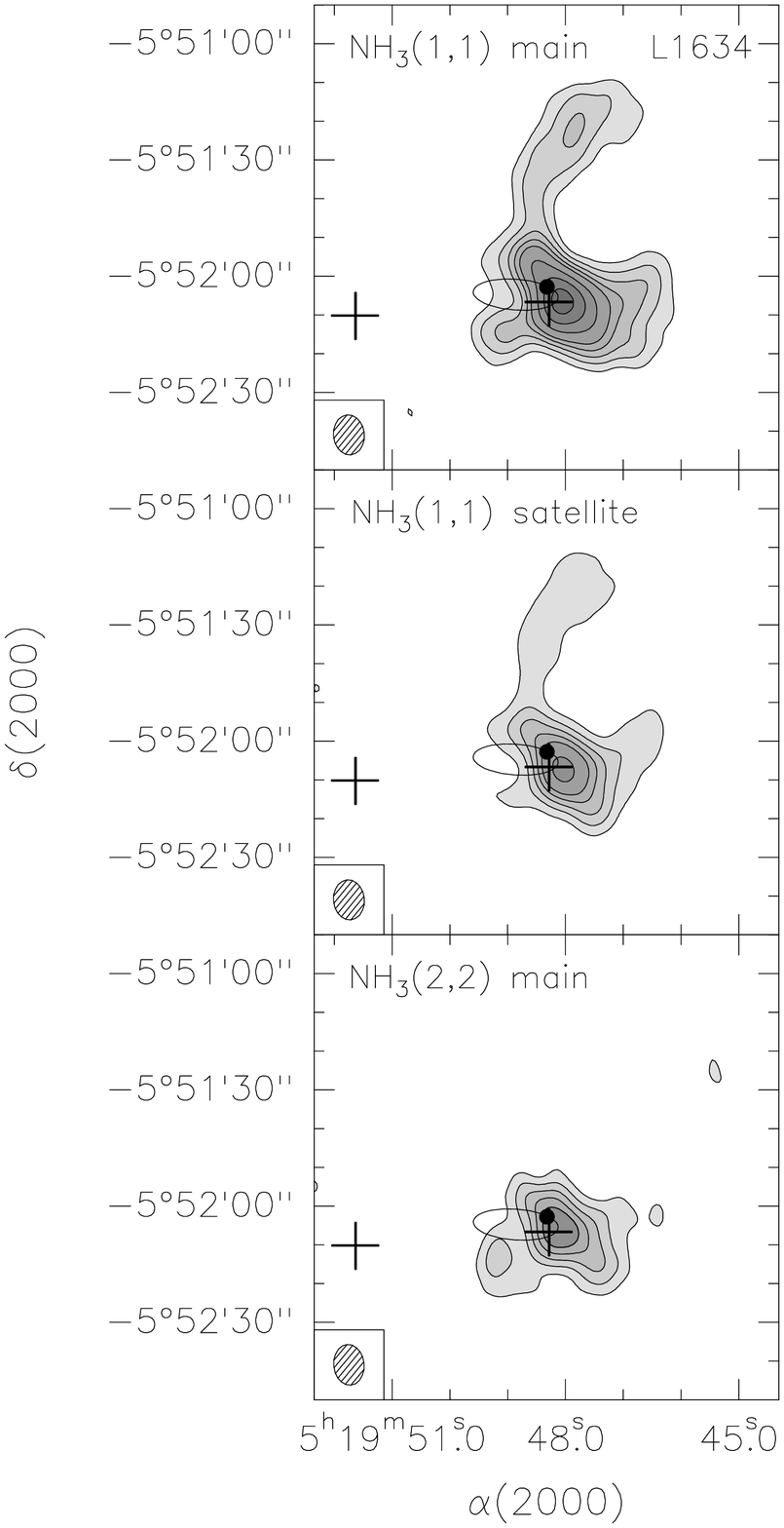}}}
\caption{({\it Top panel}): \nho\ emission averaged over the main line for a velocity
interval (7, 9)~\kms\ towards L1634. ({\it Middle panel}): \nho\ 
emission averaged over the two inner satellite lines for a velocity interval ($7$, $9$)~\kms. ({\it Bottom
panel}): \nhd\  emission averaged over the main line for a velocity
interval (7.5, 8.4)~\kms. The systemic velocity of L1634 is $8$~\kms. 
Contour levels are 3, 6, 9, 12, 15, 21, 27, and 33 times $\sigma$, where $\sigma$
is 3.5~mJy\,beam$^{-1}$ ({\it top} and {\it middle panels}) and 2.5~mJy\,beam$^{-1}$ ({\it bottom panel}). The synthesized beam is drawn in the bottom left 
corner. The crosses mark the submillimeter continuum positions of
IRAS~05173$-$0555 ({\it right cross}) and IRS~7 ({\it left cross}), and the black dot shows the position
of the 3.6~cm source associated with \iras\  (Beltr\'an et al.~\cite{beltran02}). The error
ellipse of IRAS~05173$-$0555 is shown.}
\label{nh3-aver} 
\end{figure}

\begin{figure}
\centerline{\includegraphics[angle=-90,width=8.9cm]{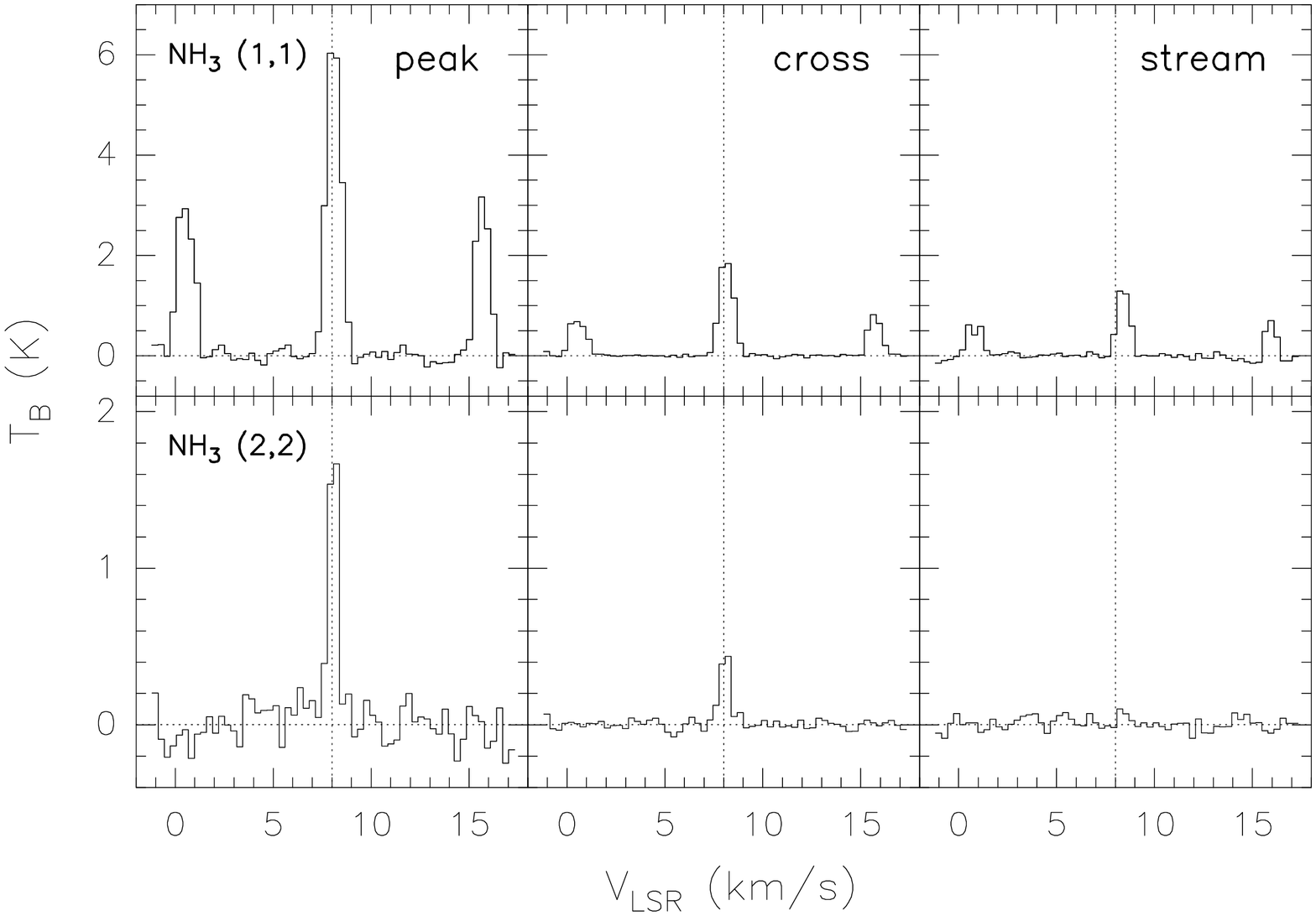}}
\caption{\nho\ ({\it top}) and \nhd\ ({\it bottom}) spectra obtained at the peak position of
the emission ({\it left panels}), and averaged over the cross-like envelope region
({\it middle panels}), and the northern stream region ({\it right panels}). 
 The dashed vertical line indicates the systemic velocity of
$8$~\kms.}
\label{lines}
\end{figure}

Figures~\ref{nh311_channel} and \ref{nh322_channel} show the velocity channel maps
respectively for the  \nho\ main and inner satellite lines emission and the \nhd\ main line emission
around the systemic velocity, \Vlsr\ $\simeq$8~\kms\ towards L1634. The emission of the \nhd\
satellite lines has not been detected. The \nho\ main and inner satellite line emission averaged
over the central channels, velocity interval (7, 9)~\kms, and the \nhd\ main line emission
averaged over the velocity interval (7.5, 8.4)~\kms\ are shown in Fig.~\ref{nh3-aver}. As can
be seen in these figures, the NH$_3$ emission is clearly associated with the embedded source
IRAS~05173$-$0555, but not with the other YSO in the region, the source IRS~7. In fact, the
\nh\ emission peaks at $\sim$$3''$ west of the submillimeter continuum position of \iras. The
submillimeter position was measured at 850~$\mu$m with the James Clerk Maxwell Telescope (JCMT), 
and have an accuracy of $\sim$2$''$.
Therefore, we cannot exclude that such a displacement is due to some pointing errors of the
observations. The \nh\ emission peak is also displaced from the position of the 3.6~cm
continuum source associated with \iras\ (Fig.~\ref{nh3-aver}). However, in this case, the
different positions could be due to the fact that the centimeter continuum emission and \nh\
are tracing different material. The \nho\ and \nhd\ spectra taken at the peak position are
shown in Fig.~\ref{lines}. Table~\ref{table_lines} lists the fitted parameters. The fits were
performed using {\it method} NH$_3$(1,1) of  the CLASS package for \nho\ and {\it method
gauss} for \nhd.

The gas of the core as traced by the \nh\ emission is resolved and shows two components clearly
distinguishable morphologically: a cross-like structure, roughly elongated in the direction of
the HH~240/241 outflow and associated with \iras, plus an arc-like stream of material elongated
towards the north.   Such a northern feature, which is not visible in \nhd, has also been
detected at submillimeter wavelengths as a stream of dust material emanating from the envelope
surrounding IRAS~05173$-$0555 (Beltr\'an et al.~\cite{beltran02}; see Fig.~\ref{850}). As can
be seen in the channel maps (Fig.~\ref{nh311_channel}), the maximum  emission of the cross-like
structure is at a velocity of $\sim$7.9~\kms, while that of the stream is at a velocity of
$\sim$8.5~\kms. In the next sections we discuss both structures in more detail.

\subsubsection{The cross-like envelope}
\label{enve}

The \nho\ emission around \iras\ is not compact but extended and elongated in the direction of
the CO outflow mapped by Lee et al.~(\cite{lee00}), which is blueshifted towards the west and
redshifted towards the east. The emission of the \nho\ main line shows a cross-like structure,
clearly visible at a velocity of 8.2 and 7.9~\kms\ (see Fig.~\ref{nh311_channel}), and in the
averaged  emission map (Fig.~\ref{nh3-aver}), with a more prominent NE-SW arm. The deconvolved
size of this cross-like structure is $\sim$$53''$  in the E-W direction and $\sim$$33''$  in
the N-S direction ($\sim$$24400$ and $\sim$$15200$~AU at the distance of the region). The
cross-like structure is more evident in the maps made with higher angular resolution (see
Fig.~\ref{line1-high-int}). Figure~\ref{line1-high-int} shows the overlay of the  \nho\
emission integrated over the main line, obtained with natural weighting and no taper applied to
the {\sl uv} data, on the blueshifted and redshifted CO~(\jtt) integrated emission, tracing the
lobes of the HH~240/241 molecular outflow. The CO~(\jtt) data are from the JCMT archive. As can
be seen in this figure, the \nho\ extended emission clearly traces the two arms of the cross,
while  towards the center of the cross \nho\ traces a more compact and flattened structure
elongated along a north-south (NW-SE) direction.  This compact structure has a deconvolved size
of $\sim$$13''\times5''$ ($\sim$$6000\times 2300$~AU) at a P.A. of $\sim$$157$\degr. This kind
of cross-like morphology has also been mapped towards other low-mass YSOs, such as L1551-IRS5
(Fuller et al.~\cite{fuller95}), B5-IRS1 (Langer et al.~\cite{langer96}), L1157 (Beltr\'an et
al.~\cite{beltran04}), HH~212 (Wiseman et al.~\cite{wiseman01}), or RNO~43 and L1228 (Arce \&
Sargent~\cite{arce06}). 

The NE-SW arm of the cross appears more prominent than the other arm. In fact, the \nho\
satellite line and \nhd\ emission traces mainly this NE-SW arm (see Figs.~\ref{nh311_channel},
\ref{nh322_channel} and \ref{nh3-aver}), although some emission from what would be the rest of
the cross-like structure is also visible at a low-emission level. As a matter of fact, the
\nhd\ emission shows an emission clump towards the southeast. This elongated NE-SW arm has a
deconvolved length of $\sim$$40''$ ($\sim$18400~AU at the distance of the region). 

The \nho\ and \nhd\ spectra integrated over the $3\,\sigma$ contour level area
($\sim$$1830$~arcsec$^2$) towards \iras, which includes all the emission from the cross-like
envelope, are shown in Fig.~\ref{lines}. Table~\ref{table_lines} lists the fitted parameters.

\subsubsection{The northern stream}

The northern arc-like stream, which is emanating from the envelope surrounding \iras, is more diffuse
than the cross-like envelope. The stream is clearly traced by the \nho\ main and satellite
lines but not by the \nhd\ line. This could be due to the physical conditions of the stream. As
can be seen in Fig.~\ref{nh311_channel}, the stream is visible for velocities redshifted with
respect to the systemic velocity of $\sim8$~\kms, which would indicate a slightly different
velocity for the stream and the envelope surrounding \iras. This stream has a deconvolved length
of $\sim$$45''$ ($\sim$$20700$~AU at the distance of the region).

The \nho\ and \nhd\ spectra integrated over the $3\,\sigma$ contour level area ($\sim$$1000$~arcsec$^2$)
towards this northern stream, which includes all the emission, are shown in Fig.~\ref{lines}.
Table~\ref{table_lines} lists the fitted parameters.

\begin{figure}
\centerline{\includegraphics[angle=-90,width=8.5cm]{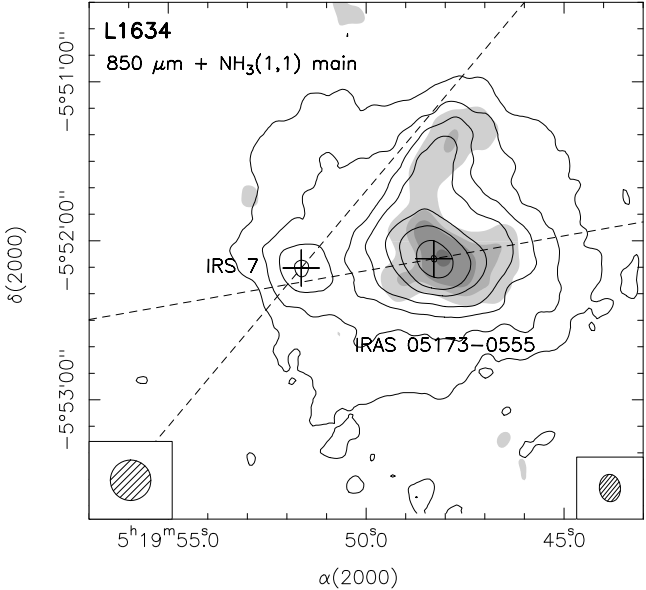}}
\caption{Overlay of the 850~$\mu$m continuum emission (Beltr\'an et al.~\cite{beltran02}) in {\it
contours} on the \nho\ emission averaged over the main line for a velocity interval 
($7$, $9$)~\kms\ in {\it gray-scale}. The crosses mark the submillimeter continuum positions of
IRAS~05173$-$0555 and IRS~7 (Beltr\'an et al.~\cite{beltran02}). Contour levels are 3, 6, 9, 12,
18, 24, 36, and 46 times 24~mJy\,beam$^{-1}$. Gray-scale levels are 3, 9, 15, and 33 times 
3.5~mJy\,beam$^{-1}$. The main beam of the JCMT 850~$\mu$m continuum observations is drawn in the
bottom left and the synthesized beam  of the VLA \nho\ observations in the bottom right. The
east-west dashed line indicates the direction of the HH~240/241 outflow, detected in H$_2$ and CO
(Davis et al. \cite{davis97}; Lee et al.~\cite{lee00}) and driven by IRAS~05173$-$0555, and 
the SE-NW dashed line indicates the direction of the H$_2$ bipolar jet (Hodapp \&
Ladd~\cite{hodapp95}) possibly powered by IRS~7.}
\label{850}
\end{figure}

\begin{figure}
\centerline{\includegraphics[angle=0,width=8.7cm]{{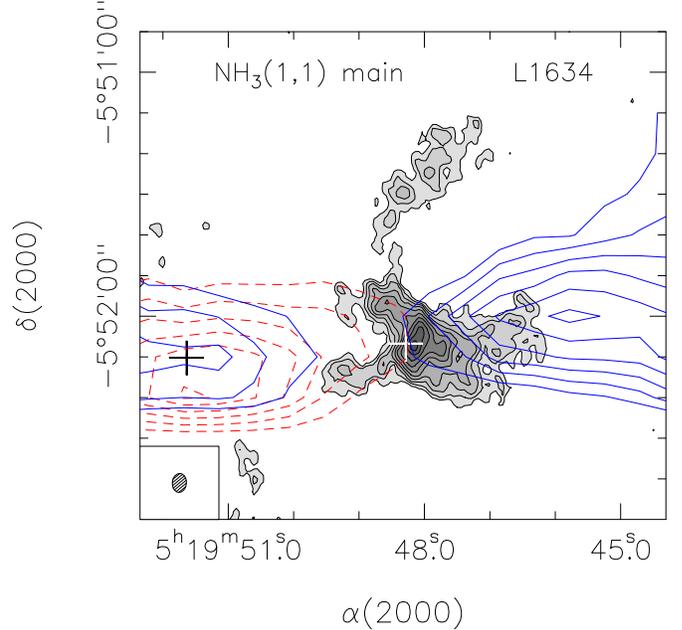}}}
\caption{\nho\ emission integrated over the main line for a velocity interval (7,9)~\kms\ towards
L1634. The synthesized beam is $4\farcs7\times3\farcs2$ at P.A.\ = $-3$\degr. The map was done 
with natural weighting and no taper applied to the {\sl uv} data. Contour levels are 3, 6, 9, 12, 15,
18, 21, 27, and 30 times $\sigma$, where $\sigma$ is 2~mJy\,beam$^{-1}$~\kms. The crosses mark the
submillimeter continuum positions of IRAS~05173$-$0555 ({\it white cross}) and IRS~7 ({\it black
cross}) (Beltr\'an et al.~\cite{beltran02}). The solid blue contours outline the blueshifted CO~(\jtt)
integrated emission in the velocity interval (2, 6.5)~\kms, and the red dashed contours the
redshifted integrated emission in the velocity interval (9.5, 12)~\kms. The
systemic velocity is 8~\kms. Contour levels are 6, 8, 10, 12, 14 and 16~K\,\kms\ ({\it blue
contours}), and 2, 4, 6, 8, 10, and 12~K\,\kms\ ({\it red contours}).}

\label{line1-high-int} 
\end{figure}

\begin{table*}
\caption{Physical parameters$^a$}
\label{table_param}
\begin{tabular}{lccccccc}
\hline
&
\multicolumn{2}{c}{Deconvolved size} &
\multicolumn{1}{c}{$T_{\rm rot}$} &
\multicolumn{1}{c}{$T_{\rm kin}$}  &
\multicolumn{1}{c}{$N({\rm H_2})$} &
\multicolumn{1}{c}{$n({\rm H_2})$}  &
\multicolumn{1}{c}{Mass}
\\
\cline{2-3}
\multicolumn{1}{c}{Region} &
\multicolumn{1}{c}{(arcsec)} &
\multicolumn{1}{c}{(pc)} &
\multicolumn{1}{c}{(K)} &
\multicolumn{1}{c}{(K)} &
\multicolumn{1}{c}{(10$^{22}$ cm$^{-2}$)} & 
\multicolumn{1}{c}{(10$^{3}$ cm$^{-3}$)} &
\multicolumn{1}{c}{($M_\odot$)} 
\\
\hline
cross-like envelope &$53\times33$ &$0.12\times 0.07$ &$12\pm1$ &13 &5.6 &7.2&11\\
northern stream &$45\times21$ &$0.10\times 0.05$ &$<9\phantom{2}$ &$<9$\phantom{2.}
&$>7.7\phantom{22}$ &$\sim$8$\phantom{1}$ &$\sim$9$\phantom{1}$ \\
\hline 
\end{tabular}

(a) See Sect.~\ref{mass} for the method of derivation of the parameters.

\end{table*}

As already mentioned, such a northern feature has also been detected at 450 and 850~$\mu$m
wavelengths as a stream of dust material emanating from the envelope surrounding IRAS~05173$-$0555
(Beltr\'an et al.~\cite{beltran02}). In fact, as can be seen in Fig.~\ref{850}, where the \nho\
emission averaged over the main line overlaps the 850~$\mu$m continuum emission (Beltr\'an et
al.~\cite{beltran02}), there is a remarkable agreement between the gas and the dust emission
towards the source IRAS~05173$-$0555 and the northern stream. 

\subsection{Mass of the gas}
\label{mass}

The physical parameters of the cross-like envelope and the northern stream are given in
Table~\ref{table_param}. The rotational temperature, $T_{\rm rot}$, was derived, following Ho
\& Townes (\cite{ho83}), from the ratio of the ammonia (1, 1) and (2, 2) column densities 
(derived taking into account the opacity of the lines and assuming the same excitation
temperature for both \nho\ and \nhd). The kinetic temperature, $T_{\rm kin}$, was estimated
from $T_{\rm rot}$, following Tafalla et al.~(\cite{tafalla04}).  The average H$_2$ column density was
calculated from Eq.~(A15) of Ungerechts et al.~(\cite{unge86}), and assuming a [H$_2$]/[NH$_3$]
abundance ratio of $10^8$ (see Anglada et al.~\cite{anglada95} for a discussion on \nh\
abundances). The H$_2$ volume density,
which is a lower limit, was
derived from the two-level model (Ho \& Townes~\cite{ho83}), assuming a filling factor $f=1$.
The excitation temperature, $T_{\rm ex}$, adopted is 5~K for the cross-like envelope, and 4.4~K
for the northern stream. It should be noted that the value of $T_{\rm ex}$ could be affected by
the assumption that $f$ is one. The \nh\ emission of the northern stream appears  more diffuse,
and therefore, such an assumption is probably correct. On the other hand, the emission of the
cross-like envelope could be more clumpy, which would imply that $f$ could be smaller than 1.
In such a case, $T_{\rm ex}$ should be higher than the value derived for the envelope. However,
the higher angular resolution map (Fig.~\ref{line1-high-int}), shows that the \nho\ emission is
not very clumpy, which suggests that $f$ for the cross-like envelope cannot be much different
of 1. The total mass of the cross-like envelope and the northern stream (see
Table~\ref{table_param}) was calculated from the average H$_2$ column density and the observed
(not deconvolved) area of each region, and assuming a mean molecular mass per H$_2$ molecule of
$2.8m_{\rm H}$, which corresponds to a 10\% Helium abundance.

As can be seen in Table~\ref{table_param}, the average value of $T_{\rm rot}$ for the northern
stream is $\sim$9~K. This suggests that the stream does not require an internal source of heating, and thus,
it could be a quiescent structure or a prestellar core with no star(s) forming inside yet. Therefore,
the different average $T_{\rm rot}$ value for the cross-like envelope and the northern stream suggests
that \nh\ could be tracing different material with different properties (see next section for a more
detailed analysis). The mass derived from the ammonia observations is $\sim$$11\, M_\odot$ for the
cross-like envelope and $\sim$$9\, M_\odot$ for the northern stream. The circumstellar mass estimated
by integrating the submillimeter continuum intensity distribution over a region of 80$''$ in diameter
around \iras\  is quite similar, $\sim$3--9~$M_\odot$, depending on the opacity law used (Beltr\'an et
al.~\cite{beltran02}). Such a region includes all the emission from the cross-like envelope and part of
the northern stream. 


\begin{figure*}
\centerline{\includegraphics[angle=-90,width=18cm]{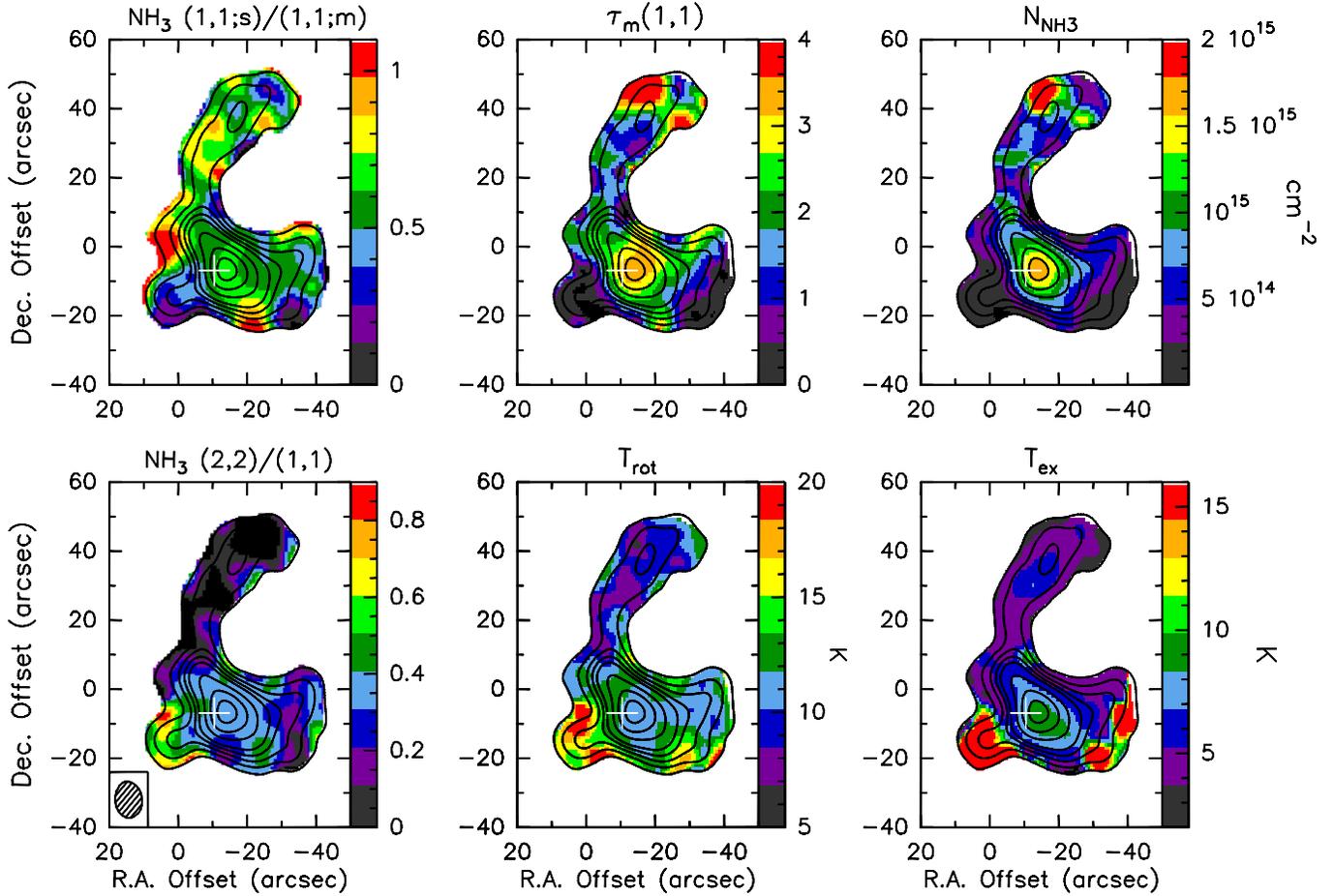}}
\caption{Overlay of the \nho\ emission averaged over the main line ({\it contours}) on the (1,1)
inner satellite/main
intensity ratio, the $\tau_{\rm m}(1,1)$, the $N$(${\rm NH_3}$), the (2,2)/(1,1) intensity
ratio, the  $T_{\rm rot}$, and 
the $T_{\rm ex}$ maps ({\it colours}). See Sect.~\ref{prop} for the method of derivation of the
parameters . The contour levels are the
same as in Fig.~\ref{nh3-aver}.  The white cross marks the submillimeter continuum position of
IRAS~05173$-$0555 (Beltr\'an et al.~\cite{beltran02}). The synthesized beam is drawn in the bottom left 
corner of the bottom left panel.} 
\label{param}
\end{figure*}

\section{Discussion}

\subsection{Properties of the emission}
\label{prop}

Figure~\ref{param} shows a map of the \nh\ (1,1) inner satellite/main intensity ratio, the optical depth
of the main quadrupole hyperfine component for the \nho\ line, $\tau_{\rm m}(1,1)$, the ammonia
total column density, $N$(NH$_3$), the (2,2)/(1,1) intensity ratio, $T_{\rm rot}$, and $T_{\rm ex}$
towards L1634.  The $\tau_{\rm m}(1,1)$, $N$(NH$_3$), $T_{\rm rot}$, and $T_{\rm ex}$ maps were
computed by fitting the \nho\ and \nhd\ spectra at each position (pixel) in the core, while the
\nh\ (1,1) satellite/main intensity ratio and (2,2)/(1,1) intensity ratio maps were obtained
directly from the emission  maps. For the sake of comparison, the \nho\ emission averaged over the
main line is also shown in Fig.~\ref{param} as the underlying contours. 

The properties of the cross-like envelope and of the northern stream are somewhat different. The
opacity $\tau_{\rm m}(1,1)$ peaks at a value of $\sim$3.5 for the cross-like envelope towards the
\iras\ position, while the maximum is $\sim$2 in the northern stream. This difference in opacity is
also visible in the \nho\ inner satellite/main intensity ratio map. The difference between the
envelope and the stream is less evident in this map. However, it is well known that the
satellite hyperfine components of \nh\ may exhibit non-LTE effects (e.g.\ Stutzki \&
Winnewisser~\cite{stutzki85}), which would influence the satellite/main intensity ratio.
 There are also some
differences in $\tau_{\rm m}(1,1)$ inside the cross-like envelope itself. The NE and NW edges
of the cross appear optically thicker than the SE and SW ones. In fact, the emission towards
the latter ones is the most optically thin in the whole region. It should be also noted the
presence of an ``optically thick spot" in the top left of the northern stream, not visible in
the \nho\ inner satellite/main intensity ratio map probably due to non-LTE effects.
 As the spot is near the edge of the map, we
checked on the quality of the spectra and the fits, which seems to be fine. The optically
thicker emission in that area translates to higher values of $N$(NH$_3$) (see Fig.\ref{param}).
 This enhancement of the opacity could be indicating the presence of an embedded source, a
possibility already suggested by Beltr\'an et al.~(\cite{beltran02}) based on the spectral
index of the dust emission towards the northern part of the stream (see their Fig.~6). We
checked the 2MASS catalog and the {\it Spitzer} MIPS and IRAC archive searching for hints of
infrared embedded protostars but found none. Therefore, this suggests that this northern
condensation does not contain any embedded source.

As can be seen in Fig.~\ref{param}, $N$(NH$_3$) is also different towards \iras\ than towards
the northern stream. $N$(NH$_3$) has a maximum value of $\sim$1.8$\pm0.2\times10^{15}$ cm$^{-2}$ 
towards \iras, and of $\sim$1.1$\pm0.2\times10^{15}$ cm$^{-2}$ towards
the stream. Again, there are some differences between the NE and NW edges of the cross-like
envelope and the SE and SW ones. $N$(NH$_3$) towards the northern ``optically thick spot" in
the stream is $\sim$3.0$\pm0.3\times10^{15}$ cm$^{-2}$. 

\begin{figure*}
\centerline{\includegraphics[angle=-90,width=14cm]{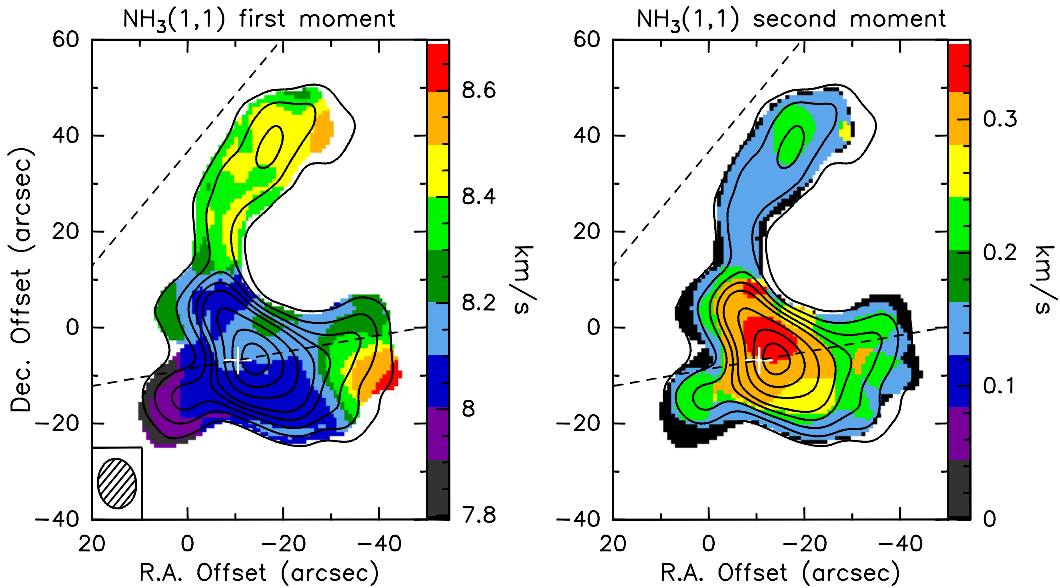}}
\caption{Overlay of the \nho\ emission averaged over the main line ({\it contours}) on the first-order
moment (intensity weighted mean \Vlsr; {\it left}), and the second-order moment 
(velocity dispersion; {\it right}) of
the \nho\ main line ({\it colours}). The contour levels are the
same as in Fig.~\ref{nh3-aver}.  The white cross marks the submillimeter continuum position of
IRAS~05173$-$0555 (Beltr\'an et al.~\cite{beltran02}). The dashed lines indicate the direction of the H$_2$ and
CO outflows in the region (Hodapp \& Ladd \cite{hodapp95}; Davis et al. \cite{davis97}; Lee et al.~\cite{lee00}).  The synthesized beam is drawn in the bottom left 
corner of the left panel.} 
\label{mom}
\end{figure*}

$T_{\rm rot}$  has a maximum value of $\sim$13~K towards the \iras\ source in the cross-like
envelope,  and of $\sim$10.5~K towards the northern stream. The difference in temperatures
between the two structures is more evident in the \nh\ (2,2)/(1,1) intensity ratio map. As can
be seen in Fig.~\ref{param}, this ratio is $\sim$0.4 towards the \iras\ position, and
$\sim$0.09 towards the stream. The \nh\ (2,2)/(1,1) ratio is very sensitive for temperatures
around 10~K. For temperatures below 10~K, the (2,2) line will become very faint, because of the
energy above ground for the (2,2) state. Therefore, the fact that the (2,2)/(1,1) ratio is so
low for the stream, clearly indicates that it has a very cold temperature ($\lesssim$10~K).  In
fact, Figs.~\ref{nh322_channel} and \ref{nh3-aver} show that the \nhd\ emission is not detected
towards the stream at a $3\,\sigma$ level. As mentioned in the previous section, such a cold
temperature suggests that there is no internal source of heating in the northern stream, and
that the structure could be a prestellar core or a quiescent cloud formed by gas that is being
dispersed. The value of $T_{\rm rot}$ estimated towards \iras\ is consistent with those found
towards other low- intermediate-mass star-forming regions (e.g.\ Jijina et
al.~\cite{jijina99}). The values of $T_{\rm rot}$ towards the cross-like envelope are
consistent with the kinetic temperature  of $\sim$14~K, derived towards \iras\ from the
CO~(\juz) brightness temperature averaged over a $2\farcm2$ region (Lee et al.~\cite{lee00};
Beltr\'an et al.~\cite{beltran02}). Note that such a large region includes emission from the
northern stream as well. In Fig.~\ref{param} a considerable enhancement of $T_{\rm rot}$ is
also visible  towards the SE (peak $T_{\rm rot}\sim$24~K) and SW (peak $T_{\rm rot}\sim$23~K)
edges of the cross. These areas correspond with the most optically thin regions, and
hence, where the best estimate of $T_{\rm rot}$ is obtained. It should be mentioned that we
checked that the signal-to-noise ratio of the spectra was high enough to properly estimate
$T_{\rm rot}$.

Finally, $T_{\rm ex}$ has a peak value of $\sim$9.3~K towards the \iras\ source in the cross-like
envelope, and of $\sim$6.5~K towards the northern stream. Therefore, $T_{\rm ex}$ also appears to be
different in both structures. The estimated $T_{\rm ex}$ in both regions is slightly lower than $T_{\rm
rot}$. As already mentioned in Sect.~\ref{mass}, this could mean that the gas is beam diluted.
Figure~\ref{param} shows that $T_{\rm ex}$ is $>12$--15~K towards the SE edge and the western
side of the cross-like envelope.


\begin{figure}
\centerline{\includegraphics[angle=0,width=8cm]{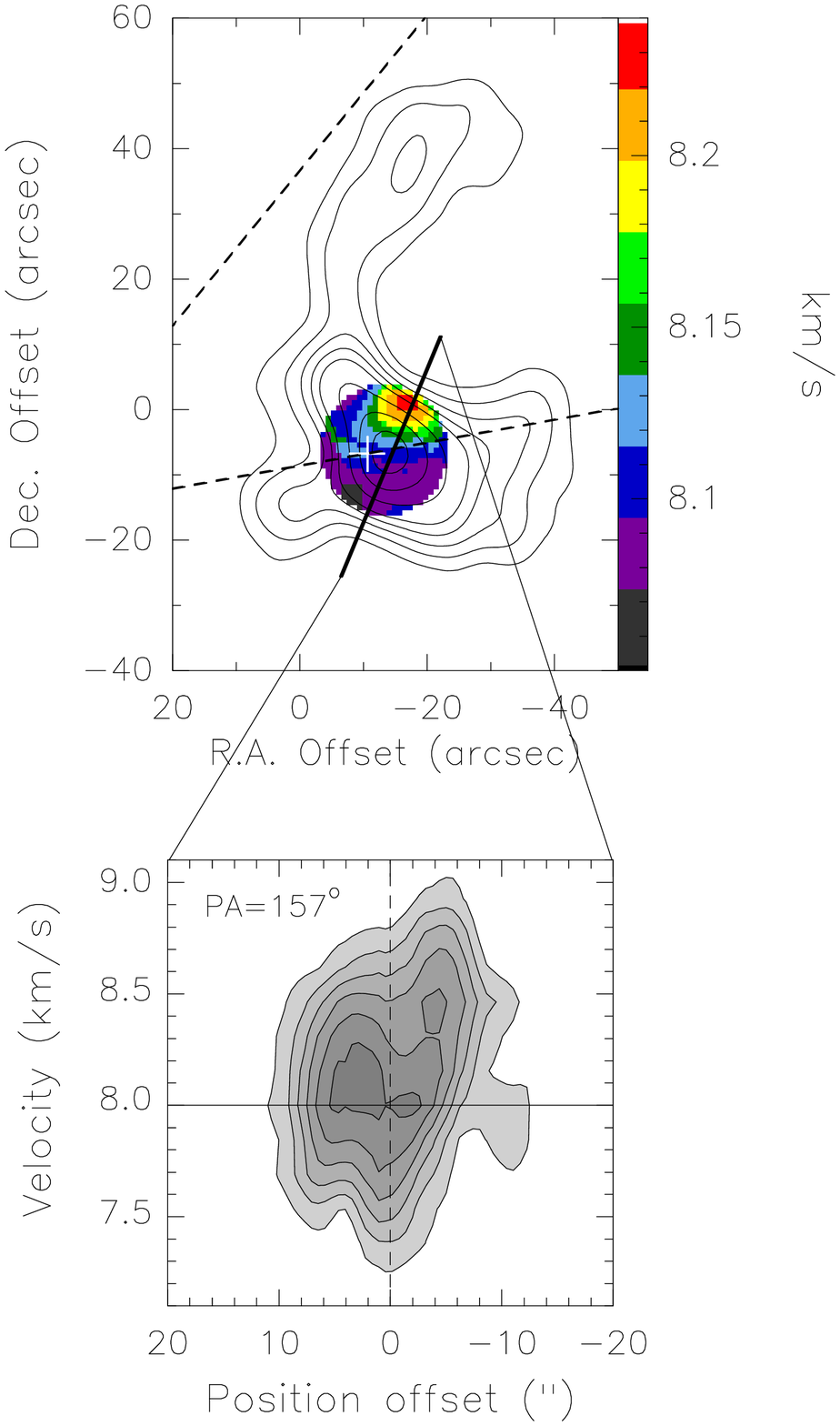}}
\caption{({\it Top}): Overlay of the \nho\ emission averaged over the main line ({\it contours}) on the first-order
moment (intensity weighted mean \Vlsr) of
the \nho\ main line in a region of 10$''$ of radius around \iras\  ({\it colours}). The contour levels are the
same as in Fig.~\ref{nh3-aver}.  The symbols are the same as in Fig.~\ref{mom}. ({\it Bottom}): PV plot of the \nho\ main line emission (map made with
natural weighting without tapering and the highest spectral resolution) along the
direction with P.A. = 157\degr, indicated by the black line in the top panel.
Contours are 3, 6, 9, 12, 15 and 18 times 3~mJy\,beam.
The
horizontal line marks the systemic velocity, \Vlsr\ = 8~\kms.} 
\label{rot}
\end{figure}

\subsection{Kinematics of the gas}
\label{kine}

Figure~\ref{mom} shows the first-order moment map (intensity weighted mean \Vlsr), and the second-order
moment (velocity dispersion) of the \nho\ main line towards L1634. As can be seen in this figure, the
velocity towards the cross-like envelope is quite different from that towards the northern stream. The
northern stream seems to have a velocity of $\sim$8.3--8.5~\kms. On the other hand, the cross-like
envelope, or at least the gas towards the \iras\ source position, seems to have a lower velocity of
$\sim$8--8.2~\kms. In addition, the velocity is different for different parts of the cross-like envelope.
In particular, the SE cross arm and the western side of the envelope are clearly at a velocity different
from the rest of the envelope. Notice that these areas are the same that show different properties, such as
$\tau_m$(1,1), $T_{\rm rot}$, or $T_{\rm ex}$. A possible explanation for this could be that these parts of
the envelope are tracing material that is being swept up or disturbed by the molecular outflow powered by
\iras. As can be seen in  Fig.~\ref{line1-high-int}, the \nho\ integrated emission, especially the eastern
side of the cross-like envelope, seems to be associated with the edges of the HH~240/241 outflow. However,
one should notice that while the cross-like envelope emission is blueshifted towards the SE and redshifted
towards the west (see Fig.~\ref{mom}), the outflow emission is blueshifted towards the west and redshifted
towards the east (see Fig.~\ref{line1-high-int}). If the velocity of the \nh\ material is caused by the
molecular outflow, dragging out the gas of the envelope, one would expect to have \Vlsr\ increasing (or
decreasing) in the same direction for the outflow and for the envelope. The high elongation of the
outflow, the spatial and velocity separation of the outflow lobes, and the relatively small velocity
width of the outflow suggest that the system is in the plane of the sky. Therefore, one can
rule out a possible face-on orientation of the envelope to explain the fact that \Vlsr\ increases (or
decreases) in
opposite directions for the outflow and the envelope. The most likely explanation for this is a combined
effect of a large scale velocity gradient and the pushing by the outflow.


The second-order moment map (Fig.~\ref{mom}) shows that the velocity dispersion increases
towards the position of the \iras\ source at the center of the cross-like envelope. As can be
seen in this map, the northern stream is hardly perturbed, as the line width of \nho\ is
$\sim$$0.15$~\kms\ across it. The maximum velocity dispersion of the northern stream is 
$\sim$$0.24$~\kms, compared to $\sim$$0.36$~\kms, which is the value measured towards the \iras\
position. This clearly suggests that the northern arc-like stream is not material that is being
entrained by the H$_2$ SE-NW outflow (Hodapp \& Ladd~\cite{hodapp95}) detected in the region.
Most likely this northern stream is part of the original cloud envelope, within which the
embedded \iras\ source has formed (see discussion in Sect.~\ref{ori_n}).


Wiseman et al.~(\cite{wiseman01}) have detected in \nh\ a rotating envelope perpendicular to the
jet in the HH~212 region, which is quite similar to L1634. Therefore, we checked for possible
rotation towards the position of the embedded \iras\ source. The presence of rotation towards
\iras\ is not evident in Fig.~\ref{mom}. However, its signature could be masked by the kinematics
of the large envelope. We searched for rotation in a region of 10$''$ in radius around the \nh\
peak, which corresponds to a radius of $\sim$5000~AU at the distance of the source.
 Figure~\ref{rot} shows the first-order moment map of the \nho\ inside this region. As can be seen
in this figure, there is a clear velocity  gradient along a well defined direction, P.A.\
$\sim$157$\degr$, coincident with the direction of the compact and flattened structure detected
with the higher angular resolution maps (see Sect.~\ref{enve} and Fig.~\ref{line1-high-int}).
Figure~\ref{rot}  also shows the  position-velocity (PV) diagram of the \nho\ main line emission,
for the maps made with natural weighting without tapering and with the highest spectral
resolution, along the direction with P.A.\ = 157\degr. The velocity gradient is of about
3.4~\kms\,pc$^{-1}$, which is slightly smaller than that of  4--5~\kms\,pc$^{-1}$ found by
Wiseman et al.~(\cite{wiseman01}) in HH~212. This velocity gradient could be indicative of
rotation of the compact core. 

The dynamical mass, $M_{\rm dyn}$, is the minimum mass for the system to be gravitationally bound.
$M_{\rm dyn}$ can be derived from the expression $M_{\rm dyn}=v_{\rm rot}^2\,R/G\,\sin^2i$, where
$v_{\rm rot}$ is the rotational velocity, $R$ is the radius of the envelope, and $i$ is the inclination
angle assumed to be 90$\degr$ for  an edge-on envelope. In this case, $R$ is $10''$ ($\sim$5000~AU),
$v_{\rm rot}$ is $\sim$0.075~\kms\ (assumed to be half of the velocity shift measured over the extent
of $\sim$10$^4$~AU), and the value of $M_{\rm dyn}$ is $\sim$0.03/$\sin^2i$~$M_\odot$. The mass of the
gas inside this region of 10$''$ in radius is $\sim$4~$M_\odot$. Regarding the stellar mass, Beltr\'an
et al.~(\cite{beltran02}) estimated a mass of $\sim$2.5~$M_\odot$ by assuming that \iras\ is on the
birth line, that its bolometric luminosity (17~$L_\odot$) is only stellar, and by using Palla \&
Stahler (\cite{palla93}) evolutionary tracks. Therefore, the fact that $M_{\rm dyn}$ is much lower
(unless the inclination angle $i$ is $\lesssim5\degr$) than the total mass (stellar+gas) of
$\sim$6.5~$M_\odot$  suggests that the motion is gravitationally bound.  The virial mass, $M_{\rm
vir}$, has been estimated from the line width of 0.53~\kms inside an area of $10''$ in radius, assuming
a spherical envelope with a power-law density distribution $\rho\propto r^{-p}$, with $p=2.0$--1.5, and
neglecting contributions from magnetic field and surface pressure (see Eq. 5 of Beltr\'an et
al.~\cite{beltran06}). $M_{\rm vir}$ is  0.8--1.0~$M_\odot$, which is also much smaller than the mass
of the gas inside this 10$''$ central region. This suggests that the material of the envelope is
sub-virial, and therefore, that the central part of the envelope might be unstable and undergoing
collapse. This is in agreement with the model of the envelope around \iras, based on dust emission,
proposed by Beltr\'an et al.~(\cite{beltran02}). These authors model the radial intensity profiles of
the extended dust emission at 850 and 450~$\mu$m by adopting power-law temperature and density
distributions, and find that the best fit is achieved with a multiple-power law that for
$R\lesssim30''$ has $\rho\propto r^{-1.5}$, that is, a free-fall collapse density distribution.
Therefore, both the estimated dynamical parameters and the dust emission indicate that the envelope
surrounding \iras\ seems to be still collapsing.

Following  Ohashi et al.~(\cite{ohashi97}), we calculated the specific angular momentum $j$,
defined as $R_{\rm rot}\times v_{\rm rot}$, where $v_{\rm rot}$ is the velocity of rotation at the
radius $R_{\rm rot}$. For the \iras\ envelope, we considered $v_{\rm rot}$$\sim$0.075~\kms\ at
$R_{\rm rot}$=$10''$ ($\sim$0.022~pc). The estimated specific angular momentum,
$j$$\sim$$1.7\times10^{-3}$~\kms\ pc, is consistent with the constant value of
$\sim$10$^{-3}$~\kms\ pc\, found by Ohashi et al.~(\cite{ohashi97})  at radii smaller than
$\sim$0.03~pc for dynamically infalling envelopes and rotationally supported disks. Therefore, as
suggested by  the estimated dynamical parameters and the dust emission (see previous paragraph),
the  inner regions of the \iras\ envelope ($R\lesssim$0.022~pc), which show conserved specific
angular momentum, are probably undergoing collapse.

One thing to notice in Fig.~\ref{rot} is that the direction of the velocity gradient is not exactly
perpendicular to the axis of the molecular outflow (P.A.\ $\sim$100\degr), which is what one expects if the
core is rotating about the axis of the outflow. Nevertheless, as can be seen in the CO maps of Fig.~3 of
Lee et al.~(\cite{lee00}), the direction of the molecular outflow is quite difficult to determine, as it
has a different position angle depending on the velocity and the distance to the powering source. In fact,
as pointed out by Davis et al.~(\cite{davis97}), the flow might be slowly changing direction (wiggling),
since the HH~240 and HH~241 knots do not lie on the same axis. Thus, the outflow could be precessing.
Therefore, although at a large scale the outflow shows a P.A.\ $\sim$100\degr, one cannot discard the
possibility that the outflow closer to the powering source has a different direction and that the velocity
gradient observed in NH$_3$ is perpendicular to the outflow axis close to the \iras\ source.

\subsection{Origin of the cross-like envelope}

As mentioned in the previous sections, the morphology of the envelope surrounding \iras\
clearly resembles a cross, that is suggestive of interaction between the outflow and the
envelope. In such a scenario, the outflow would be sweeping up envelope material, in this case
traced in \nh, which would
outline the beginning of the outflow conical lobes; that is, the walls of the cavity excavated
by the outflow. This scenario seems to be supported by the properties and the kinematics of the
gas in the cross-like envelope. In fact, the SE and SW edges of the cross have the most
optically thin emission, which would be consistent with more diffuse envelope material (located
farther from the embedded source) being dragged out by the outflow.  $T_{\rm rot}$ is also
considerably enhanced towards the SE and SW edges as compared to the temperature  towards the
\iras\ position (see Sect. \ref{prop}), which also suggests interaction between the outflow and
the envelope. Note that as seen in the previous section, the central part of the envelope might
be unstable and undergoing collapse. Therefore, this suggests that while the envelope material
located farther from the embedded \iras\ source could be being dispersed by the molecular
outflow, the central part of the envelope could still be collapsing.

\subsection{Origin of the northern stream}
\label{ori_n}

 The properties and kinematics of the gas towards the northern arc-like stream suggest that
it might be tracing material different from that of the cross-like envelope, or material that
is being perturbed in a different way (see Sects.~\ref{prop} and \ref{kine}). The velocity
dispersion and the low $T_{\rm rot}$ and $T_{\rm kin}$  of the gas rule out the possibility
that the northern stream is
material of the envelope surrounding \iras\ that has been entrained by the SE-NW H$_2$ flow.


On the other hand, the stream could be part of the original cloud envelope. The spectral index map of
the dust emission (see Fig.~6 of Beltr\'an et al.~\cite{beltran02}) has similar values for the cross-like
envelope and the northern stream, which suggests that the material in the stream has the same dust
properties as the material surrounding the \iras\ source.  In addition, the shape of the \nh\ and the dust
emission (see Fig.~\ref{850}) resembles that of the cloud at a larger scale as traced in CO, C$^{18}$O, and
HCO$^+$ (De Vries et al.~\cite{devries02}). According to these authors, the cloud seems to be shaped by the
ionizing radiation coming from Barnard's Loop, which could also be responsible for the morphology of the
\nh\ and dust emission, and in particular, for the arc-like northern stream. Therefore, the stream could be a
remnant of the star-formation process in L1634, which might be in a prestellar phase.  Or, it could be a
quiescent core, as suggested by the cold $T_{\rm rot}$, that might never form stars and that will be
finally dispersed.




\section{Conclusions}

We studied with the VLA at 1.3~cm, the ammonia emission towards the core of L1634, a region
that contains two embedded YSOs, IRAS~05173$-$0555 and IRS~7,  powering two outflows.

The \nh\ emission has clearly been detected towards the young stellar object IRAS~05173$-$0555
but not towards IRS~7. The gas of the core as traced by the \nh\ emission is resolved and shows
two components clearly distinguishable morphologically: a cross-like structure, roughly
elongated in the direction of the HH~240/241 outflow and associated with \iras, plus an
arc-like stream elongated towards the north.  

The \nho\ emission clearly traces the two arms of the cross-like envelope.  The deconvolved
size of this cross-like structure is $\sim$$53''\times33''$ ($\sim$$24400\times15200$~AU). The
mass derived from the ammonia observations is $\sim$$11\, M_\odot$, and $T_{\rm
kin}\simeq$13~K. The emission of the cross-like structure is maximum at a velocity of
$\sim$7.9~\kms. The properties of the gas vary across the envelope, with two parts of the
cross-like envelope clearly differentiated from the rest, the SE and the SW edges of the
cross. The morphology, properties and kinematics of the gas suggest that the origin of the cross-like
morphology could be the interaction between the outflow and the envelope.  The outflow could
be sweeping up envelope material, which would outline the beginning of the outflow conical
lobes.

A more compact and flattened structure, with a deconvolved size of $\sim$$13''\times5''$
($\sim$$6000\times 2300$~AU), has been detected towards the center of the cross-like envelope and
orientated perpendicular to it, P.A. $\sim$$157$\degr. This structure could be undergoing rotation
about the axis of the outflow, as suggested by the velocity gradient detected and the PV plot
along the gradient direction. The fact that the virial mass is much
smaller than the mass of the gas suggests that this flattened envelope might be unstable and
undergoing collapse. The collapse is also suggested by the modeling of the dust emission and the
the specific angular momentum of the inner region of the \iras\ envelope. Therefore, this suggests that while the envelope material located farther
from the embedded \iras\ source could be being dispersed by the molecular outflow, the central
part of the envelope could still be collapsing.

The northern stream is clearly traced by the \nho\ main and satellite lines but not by the
\nhd\ line. This stream has a deconvolved length of $\sim$$45''$ ($\sim20700$~AU). The mass
derived from the ammonia observations is  $\sim$$9\, M_\odot$, and $T_{\rm kin}<9$~K. The
stream is visible for velocities redshifted with respect to the systemic velocity of
$\sim8$~\kms, which would indicate a velocity for the stream slightly different from that of
the envelope surrounding \iras. The velocity dispersion of the stream is very small indicating
that the gas is hardly perturbed. This arc-like stream of material, which has properties
different from those of the cross-like envelope, could be a remnant of the star-formation
process in L1634, and might be in a prestellar phase. Or, it could be a quiescent core, as
suggested by the cold $T_{\rm rot}$, that might never form stars and that will be finally
dispersed.

\begin{acknowledgements}

 We acknowledge the anonymous referee for his/her useful comments. MTB and RE are supported by MEC grant
AYA2005-08523-C03. This research has made use of the NASA/IPAC Infrared Science Archive, which is operated by
the Jet Propulsion Laboratory, California Institute of Technology, under contract with the National
Aeronautics and Space Administration (NASA).

\end{acknowledgements}


%

%
%


\begin{thebibliography}{}

\bibitem[2006]{arce06}
Arce, H.\ G., \& Sargent, A.\ I.\ 2006, ApJ, 646, 1070


\bibitem[1987]{adams87} 
Adams, F.\ C., Lada, C.\ J., \& Shu, F.\ H.\ 1987, ApJ, 312, 788


\bibitem[1995]{anglada95} 
Anglada, G., Estalella, R., Mauersberger, R., Torrelles, J.\ M., et al.\  1995, ApJ, 443,
682




\bibitem[2002]{beltran02} 
Beltr\'an, M.\ T., Estalella, R., Ho, P.\ T.\ P., Calvet, N., Anglada, G., \& Sep\'ulveda, I.\ 2002, ApJ,
565, 1069

\bibitem[2006]{beltran06} 
Beltr\'an, M.\ T., Girart, J.\ M., \& Estalella, R.\ 2006, A\&A, 457, 865

\bibitem[2004]{beltran04} 
Beltr\'an, M.\ T., Gueth, F., Guilloteau, S., \& Dutrey, A.\ 2004, A\&A, 416, 631



\bibitem[1993]{bohigas93} 
Bohigas, J., Persi, P., \& Tapia, M.\ 1993, A\&A, 267, 168



%
\bibitem[1997]{chini97} 
Chini, R., Reipurth, B., Sievers, A., Ward-Thompson, D.,
Haslam, C.\ G.\ T., Kreysa, E., \& Lemke, R.\ 1997, A\&A, 325, 542  



\bibitem[1997]{davis97} 
Davis, C.\ J., Ray, T.\ P., Eisl\"offel, J., \& Corcoran, D.\ 1997, A\&A, 324, 263

\bibitem[2002]{devries02}
De Vries, C.\ H., Narayanan, G., Snell, R.\ L.\ 2002, ApJ, 577, 798

%

%


\bibitem[2005]{froebrich05}
Froebrich, D.\ 2005, ApJSS, 156, 169

\bibitem[1995]{fuller95}
Fuller, G.\ A., Ladd, E.\ F., Padman, R., Myers, P.\ C., \& Adams, F.\ C.\ 1995, ApJ, 454, 862 


\bibitem[1983]{ho83}
Ho, P.\ T.\ P., \& Townes, C.\ H.\ 1983, ARA\&A, 21, 239

\bibitem[1995]{hodapp95} 
Hodapp, K.-W., \& Ladd, E.\ F.\ 1995, ApJ, 453, 715

%
%
%

\bibitem[1999]{jijina99}
Jijina, J., Myers, P.\ C., \& Adams, F.\ C.\ 1999, ApJSS, 125, 161



%

\bibitem[1996]{langer96}
Langer, W.\ D., Velusamy, T., \& Xie, T.\ 1996, ApJ, 468, L41

\bibitem[1969]{larson69} 
 Larson, R.\ B.\ 1969, \mnras, 145, 271

%
\bibitem[2000]{lee00} 
Lee, C.-F., Mundy, L.\ G., Reipurth, B., Ostriker, E.\ C., \& Stone, J.\ M.\ 2000, ApJ, 542, 925


\bibitem[1986]{maddalena86} 
Maddalena, R.\ J., Morris, M., Moscowitz, J., \& Thaddeus, P. 1986, \apj, 303, 375

%
%

\bibitem[2008]{morgan08}
Morgan, L.\ K., Thompson, M.\ A., Urquhart, J.\ S., \& White, G.\ J.\ 2008, A\&A, 477, 557

\bibitem[2002]{nisini02}
Nisini, B., Caratti o Garatti, A., Giannini, T., \& Lorenzetti, D.\ 2002, A\&A, 393, 1035


\bibitem[2004]{oconnell04}
O'Connell, B., Smith, M.\ D., Davis, C.\ J., Hodapp, K.\ W., Khanzadyan, T., \& Ray, T.\ 2004,
A\&A, 419, 975


\bibitem[1997]{ohashi97}
Ohashi, N., Hayashi, M., Ho, P.\ T.\ P., Momose, M.\ T.\ et al.\ 1997, ApJ, 488, 317


%
%
\bibitem[1993]{palla93} 
Palla, F., \& Stahler, S.\ W.\ 1993, ApJ, 418, 414
%

\bibitem[1993]{reipurth93}
 Reipurth, B., Chini, R., Kr\"ugel, E., \& Sievers, A.\ 1993, A\&A, 273, 221

%
%
%
%
%

\bibitem[2008]{saigo08}
Saigo, K., Tomisaka, K., \& Matsumoto, T.\ 2008, ApJ, 674, 997

\bibitem[1987]{shu87} 
Shu, F.\ H., Adams, F.\ C.\, \& Lizano, S.\ 1987, ARA\&A, 25, 23

%
%
%
%

\bibitem[1985]{stutzki85}
Stutzki, J., \& Winnewisser, G.\ 1985, A\&A, 144, 13

\bibitem[1991]{sugitani91}
Sugitani, K., Fukui, Y., \& ogura, K.\ 1991, ApJS, 77, 59

\bibitem[2004]{tafalla04}
Tafalla, M., Myers, P.\ C., Caselli, P., \& Walmsley, C.\ M.\ 2004, A\&A, 416, 191

%
%
%

\bibitem[1986]{unge86}
Ungerechts, H., Walmsley, C.\ M., \& Winnewisser, G.\ 1986, A\&A, 157, 207


\bibitem[2001]{wiseman01}
Wiseman, J., Wootten, A., Zinnecker, H., \& McCaughrean. M.\ 2001, ApJ, 550, L87

\end{thebibliography}
\end{document}